**RESEARCH ARTICLE**

JASIST WILEY

# Open is not forever: A study of vanished open access journals

**Mikael Laakso[1]**  | **Lisa Matthias[2]** | **Najko Jahn[3]**

[1]Information Systems Science, Hanken School of Economics, Helsinki, Finland

[2]Department of Political Science, John F. Kennedy Institute, Freie Universität Berlin, Berlin, Germany

[3]Göttingen State and University Library, University of Göttingen, Göttingen, Germany

**Correspondence**
Mikael Laakso, Information Systems Science, Hanken School of Economics, Arkadiankatu 22, 00100 Helsinki, Finland.
Email: mikael.laakso@hanken.fi

**Abstract**
The preservation of the scholarly record has been a point of concern since the beginning of knowledge production. With print publications, the responsibility rested primarily with librarians, but the shift toward digital publishing and, in particular, the introduction of open access (OA) have caused ambiguity and complexity. Consequently, the long-term accessibility of journals is not always guaranteed, and they can even disappear from the web completely. The focus of this exploratory study is on the phenomenon of vanished journals, something that has not been carried out before. For the analysis, we consulted several major bibliographic indexes, such as Scopus, Ulrichsweb, and the Directory of Open Access Journals, and traced the journals through the Internet Archive's Wayback Machine. We found 174 OA journals that, through lack of comprehensive and open archives, vanished from the web between 2000 and 2019, spanning all major research disciplines and geographic regions of the world. Our results raise vital concern for the integrity of the scholarly record and highlight the urgency to take collaborative action to ensure continued access and prevent the loss of more scholarly knowledge. We encourage those interested in the phenomenon of vanished journals to use the public dataset for their own research.

## 1 | INTRODUCTION

The preservation of the scholarly record requires sustained and direct action, which begins with the question of responsibility. *Preservation* refers to a set of activities to ensure the long-term accessibility and usability of original material, such as environmental control, disaster planning, storage and handling, digitization, and maintenance of digitally stored material (American Library Association, 2008; Northeast Document Conservation Center, 2015; UNESCO/UBC Vancouver Declaration, 2012). Library collections of printed academic journals and books secure long-term access through physical copies, but the shift from analog to digital gave rise to uncertainty as to who is responsible for preserving scholarly literature in electronic formats—publishers, libraries, authors, or academic institutions (Day, 1998; Fenton, 2006; Johnson, Watkinson, & Mabe, 2018; Meddings, 2011; Moulaison & Million, 2015; Science Europe, 2018; Waters, 2005). This ambiguity can be dangerous since electronic resources are vulnerable to various threats, such as hardware or software failure, natural disasters, or economic failure. If there is no general agreement whose responsibility it is to preserve electronic resources, no one will be responsible, and we risk losing large parts of the scholarly record due to inaction. Exactly how much digital journal content has already been lost is unknown since the data needed to assess the gravity of the situation is not collected anywhere, which also complicates assessing the risk of journals vanishing in the future. The dynamic nature of the scholarly publishing landscape adds to the difficulty of such data collection efforts—new journals launch while others cease publication, some change their name or publisher, while others flip to open access (OA) or reverse-flip to a subscription model (Laakso, Solomon, & Björk, 2016; Matthias, Jahn, & Laakso, 2019). Commercial indexing services could be a starting point for such endeavors. However, neither Web of Science nor Scopus provide a comprehensive or representative view of the global journal landscape as their indexing strategies introduce linguistic, geographical, and disciplinary biases (Elsevier, n.d.; Clarivate Analytics, n.d.; Mongeon & Paul-Hus, 2016). Compared to Web of Science or Scopus, Ulrichsweb provides more comprehensive serial coverage, yet the classification of active and inactive journals is not always accurate (Mongeon & Paul-Hus, 2016). The difficulties of tracing inactive journals prevail even when drawing on multiple key data sources and trying to connect the different data points, such as consulting the ISSN Portal and DOI registration agencies like Crossref or DataCite.

The limited data availability and the current gap in the literature do not mean that this is a trivial issue or that the scholarly community has successfully solved the issue of digital preservation. While all digital journals are subject to the same threats, OA journals face unique challenges. Efforts around preservation and continued access are often aimed at securing postcancellation access to subscription journals—content the library has already paid for. The same financial incentives do not exist when journals are freely available. Furthermore, unlike closed-access journals, which commonly secure funds through subscriptions, OA journals rely on alternative funding sources, such as article processing charges (APCs) or sponsorships, to finance their publishing activities (Björk, Shen, & Laakso, 2016; Morrison, 2016). Especially small-scale and APC-free journals might have limited financial resources and, as a way to keep operating costs low, might opt for lightweight technical solutions, such as university websites and servers or content management systems like WordPress (Adema & Stone, 2017; Brown, 2013). However, these options do not protect against technical instabilities, and if the journals cannot afford to enroll in preservation schemes, long-term access to their websites cannot be ensured (Lightfoot, 2016; Marchitelli, Galimberti, Bollini, & Mitchell, 2017). OA journals are enrolled in preservation schemes at an alarmingly low rate, with 4 in 10 journals indexed in the DOAJ reporting enrollment in at least one preservation or archiving scheme (5,881 out of 14,068 journals; DOAJ, 2019).

These numbers and the prevailing uncertainty surrounding the persistence of OA journals prompted us to take a closer look at what is no longer there. To our knowledge, there have been no studies analyzing scholarly journals that were openly available on the web at one point but have since disappeared to assess the scope and extent to which OA journals are vanishing. In particular, we sought to determine how many OA journals we have lost comprehensive access to for lack of preservation arrangements. Additionally, we examine the background of these vanished journals to learn about their publishing lifespan as well as their geographical and disciplinary distribution.

## 2 | LITERATURE REVIEW

Although scholars have emphasized the urgent need for action to preserve the scholarly record (cf. Barnes, 1997; Case, 2016; Hodgson, 2014; Waters, 2005), the enrollment in preservation schemes has progressed only slowly. In order to monitor such arrangements, Jisc in the United Kingdom funded a project to Pilot an E-Journal Preservation Registry Services (PEPRS) that eventually became the Keepers Registry, which currently monitors 13 preservation schemes (Burnhill, 2009, 2013; Burnhill & Guy, 2010). Shortly after the initial launch in 2011, the Keepers recorded 16,558 serial titles with preservation agreements in place. By 2015, that number had increased to 27,463 and had reached 43,647 by 2019 (Burnhill & Otty, 2015; The Keepers Registry, 2018). During the same time, however, the total number of serial titles had increased by almost threefold, from 97,563 titles in 2011 to 269,868 in 2019 (ISSN International Centre, 2020). So, while the total number of titles enrolled in preservation schemes almost tripled, the proportion of preserved titles compared to all existing titles has stayed the same at around 16%. It is worth highlighting that ISSNs are assigned to various content formats and not just limited to scholarly journals, which makes more focused comparative analyses difficult.

To contextualize our work, we introduce the four most prominent preservation initiatives relevant to scholarly journals (see Table 1) and present a brief review of the discussion around the roles and responsibilities of different actors in the maintenance and support of these services (Galyani Moghaddam, 2008; Kenney, Entlich, Hirtle, McGovern, & Buckley, 2006; Mering, 2015).

### 2.1 | Preservation services: Many options, slow uptake

The first initiative to preserve digital scholarly journal content, LOCKSS (Lots of Copies Keep Stuff Safe), was launched by Stanford University Libraries in 1999 (Reich & Rosenthal, 2001). The development of LOCKSS has been closely aligned with the needs of the library community and emulates the traditional paper-based preservation system where several copies of individual pieces of content are distributed around the world so that the material remains accessible should a copy be lost or destroyed. In a similar vein, LOCKSS ensures perpetual, real-time access to digital material by operating through a decentralized and distributed peer-to-peer network. With permission from the publisher, individual libraries, or peers, build local archives of their OA and subscription collections. At the same time, each library is also connected to several other participating libraries to compare identical copies and keep them intact. Should a copy within the network be damaged or lost, it is repaired or replaced through another library's copy. Easing possible tensions with publishers, copyright restrictions for preserved content remain in place as long as the file is available through the publisher, so that users can only access content from their own library's collection. Once preserved content becomes unavailable from the publisher, LOCKSS lifts these restrictions and provides access for anyone. Since the initial launch more than two decades ago, LOCKSS has become the most widely used preservation scheme among academic institutions, which fund the preservation network through membership fees.

**TABLE 1** Comparison of prominent international preservation services (Public Knowledge Project, n.d.; Portico, n.d.; CLOCKSS Archive, LOCKSS Program, & Portico, 2019)

| | LOCKSS | Portico | CLOCKSS | PKP PN |
|---|---|---|---|---|
| Launched in | 1999 | 2005 | 2006 | 2016 |
| Governance | Stanford libraries | Ithaka, nonprofit; advisory committee of Libraries and publishers | Freestanding nonprofit; board of directors Comprised of 12 libraries And 12 publishers | PKP (Stanford University and Simon Fraser University Library) |
| Costs (in USD, per year) | Libraries: $2,600–13,200 Publishers: Free | Libraries: $1,600–25,500 Publishers: $250–82,000 | Libraries: $485–16,140 Publishers: $242–28,500 | Free |
| Open Source | Yes | Yes | Yes | Yes |
| Preservation method | Automatic: web harvest or OAI-PMH | Manual: file transfer | Automatic and manual: web harvest, file transfer, or OAI-PMH | Automatic |
| What is preserved? | All scholarly journal content from participating libraries | All scholarly journal content from participating libraries and publishers | All scholarly journal content | Content from participating OJS-based journals |
| Journal content | 11,000 journal titles | 33,803 journal titles | 26,000 journal titles | 1,638 participating, OJS-based journals |
| Access to preserved content | Real-time backup for temporary downtimes: participating institutions After trigger event: everyone | After trigger event Closed access content: participating institutions Open access content: everyone | After trigger event: everyone | After trigger event: everyone |

The open-source licensing of LOCKSS has enabled several other preservation services to use the software to set up Private LOCKSS Networks (PLNs; Reich & Rosenthal, 2009), such as Controlled LOCKSS (CLOCKSS) or the Public Knowledge Project Preservation Network (PKP PN). CLOCKSS, which currently is the PLN with the broadest content coverage, was founded in 2006 as a cooperation between research libraries and academic publishers to ensure access to digital material. Participating publishers allow CLOCKSS to preserve their content continuously, so it remains accessible even after it becomes unavailable through the publisher, or "triggered." As Table 1 outlines, CLOCKSS covers 26,000 titles of which only 64 journals (i.e., 0.02%) have been triggered so far (CLOCKSS, 2020a). In contrast to LOCKSS, libraries also archive content they do not subscribe to (CLOCKSS, n.d.-a). This content only becomes available after a trigger event and the approval from the Board of Directors (CLOCKSS, n.d.-b) and is assigned "open" Creative Commons licenses. A collaboration between the DOAJ and CLOCKSS that could have improved the long-term preservation coverage of DOAJ journals did not come to fruition (Mitchell & Dyck, 2018). PKP PN functions in a similar way and ensures that preserved content remains accessible to everyone once it has been triggered. The most striking difference is that PKP PN presents an entirely free solution for PKP's Open Journal Systems (OJS) journals, which might otherwise not be able to afford membership fees to other preservation schemes (Sprout & Jordan, 2018). PKP PN has faced some functionality challenges in the last year but has since started resolving these issues through enhancing internal project management (PKP News, 2020).

Similar to CLOCKSS, Portico is jointly governed by an advisory committee consisting of librarians and publishers. Ithaka, the organization behind JSTOR, launched the preservation scheme in 2005 (Fenton, 2006). As with CLOCKSS and PKP PN, Portico only provides access to triggered content, with the difference that once paywalled content only becomes accessible to libraries participating in Portico, yet regardless of whether these libraries had a subscription to the triggered title or not. Until recently, such paywalls have also applied to triggered OA content, but Portico has since changed its access policy so that OA titles remain openly available for everyone (Wittenberg, Glasser, Kirchhoff, Morrissey, & Orphan, 2018). Portico covers 33,803 journals, of which 138 titles (i.e., 0.04%) have been triggered to date (Portico, 2020). Besides the higher release threshold and unlike LOCKSS-based systems, Portico operates as a proprietary, centralized archive and does not require libraries to maintain their own servers. However, outsourcing maintenance tasks to Portico also involves significantly higher annual costs that not all libraries might be able to afford.

Each of the initiatives introduced above comes with its own benefits and drawbacks, so it is perhaps not surprising to see journals use multiple preservation services in parallel as a 2011 study found substantial overlap in the coverage of three of the services—CLOCKSS, LOCKSS, and Portico (Seadle, 2011). In particular, the study found that 62% of Portico's holdings are also preserved by either CLOCKSS or LOCKSS and account for more than two-thirds all preserved content in CLOCKSS and LOCKSS. The same study found that while all three services included publishers of all sizes, large publishers more frequently use CLOCKSS and Portico, whereas small publishers turn to LOCKSS (ibid.). The author notes, "this means that, in the real world at present, the only archiving system that genuinely protects endangered content is LOCKSS—if only because it is the only system that they can afford" (Seadle, 2011, p. 194).

## 2.2 | Roles and responsibilities

Although the necessary infrastructure exists, at least to some extent, questions as to what content to preserve and who should be responsible for its preservation remain unresolved. Current practices for selecting content for preservation can disadvantage OA content since aspects such as journal impact factors or the invested cost for content acquisition often drive such decisions (Choi & Park, 2007; Wittenberg et al., 2018). Especially in the case of small and independent OA journals, which face financial and technical barriers to preservation arrangements (Regan, 2016), it seems that the opposite approach for content selection is needed—one that also includes the most vulnerable journals instead of prioritizing prestige. Indeed, preserving the "long tail" of scholarly literature might be one of the most pressing challenges the scholarly community is facing.

The most prominent actors in the space of digital preservation are publishers and libraries, who often come together to cooperate but often do so from different perspectives. As Fenton (2006, pp. 82–83) writes, "publishers are understandably eager to ensure that access to archived literature does not reduce the value of their current product offerings" and "while preservation may not be mission critical for publishers, it is at the heart of the work of many libraries." Still, a study from 2017 indicates that a strong commitment to preservation in the form of institutional policies is not common practice for libraries yet (Dressler, 2017). Surveying the 124 members of the Association of Research Libraries, Dressler (2017) found that only 32 libraries had implemented such policies, and an additional 23 were in the works. While most policies mentioned LOCKSS (9 out of 32 policies), libraries commit to content preservation in a variety of ways, for example, by choosing different preservation schemes or relying on institutional repositories as a means of archiving (Adema, Stone, & Keene, 2017). A critical challenge for libraries is the limited financial resources available to them. Since dedicated preservation funds are rare, preservation efforts often need to draw on funds for other core library activities, which makes it challenging to invest in institutional repositories, participate in initiatives like CLOCKSS, or collaborate with new types of actors like Google (Bogdanski, 2006). Such investment decisions are further complicated by the absence of a universal solution or a clear market leader (Kenney et al., 2006). National or consortial collaboration could be a way forward in tackling these challenges as this would enable libraries to act as a collective instead of fending for themselves (Barnes, 1997; Burnhill & Otty, 2015). Other studies point toward closer collaboration between libraries and individual scholars, underlining the importance of addressing the issue of preservation as early as possible in the publishing process (Harkema & Nelson, 2013; Moulaison & Million, 2015). Establishing preservation as an integral part of the publishing process would have the benefit of coordinating efforts and creating a sense of shared responsibility to act more effectively.

Literature directly related to vanishing journals or articles is surprisingly scarce. One of the few studies that focuses on this issue is Lightfoot from 2016 that found the websites of 2% of all DOAJ-indexed journals ($n = 9,073$) to be no longer available. However, the study has three main limitations. First, Lightfoot only checked for access to the journal websites but not the published content. Second, relying on the URL currently listed in the DOAJ is problematic in cases where the journal's URL has changed, and no redirect is in place. Finally, and perhaps most notably, the sample was biased against vanished journals since the DOAJ only indexes journals that are actively publishing. Similarly to Lightfoot, Marchitelli et al. (2017) assessed changes in the availability of OA journals over time and identified 122 journals that had been removed from the DOAJ because the reported journal website became unavailable.

This is where earlier studies have stopped their inquiries but where we have decided to begin ours. In particular, we seek to answer the following research questions:

1. How many OA journals have vanished from the web?
2. When did the OA journals vanish from the web?
3. What are the characteristics of vanished OA journals?

## 3 | MATERIAL AND METHODS

### 3.1 | Identifying vanished journals

The first challenge we faced was to identify vanished journals. We define a "vanished" OA journal as a journal that published at least one volume as immediate OA after which production ceased, and the journal, together with the published full-text documents, disappeared from the web. We note that in some cases, individual issues of the vanished journals might still exist on the web, through subscription services, such as EBSCO or Proquest, or as paper copies in a library—the latter particularly concerns print subscription journals that were already active before adopting digital and OA formats. However, the critical aspect in each of these scenarios is that from the moment the journal vanished from the web, access was no longer open and comprehensive. Hence, we consider journals as vanished when we find <50% of their volumes to be openly available at the time of data collection, and even if it is possible that individual journal issues are available for on-premise use or through commercial subscription services. We tried to find openly accessible versions of the journals several times during the last 12 months (September 2019–September 2020); our latest attempt was between September 1st and 3rd, 2020.

We focus on journals instead of articles for methodological reasons. For an article-level analysis, published content could be identified through persistent identifiers, such as DOIs or ORCiD iDs (Klein & Van de Sompel, 2017; Van de Sompel, Rosenthal, & Nelson, 2016). However, since neither is used universally yet (Boudry & Chartron, 2017; Gorraiz, Melero-Fuentes, Gumpenberger, & Valderrama-Zurián, 2016), a considerable amount of vanished content would likely go undetected, and so we found this approach unsuitable for the current paper. A journal-level approach, on the other hand, is challenging because no single data source exists that tracks the availability and accessibility of journals over time. Large indexes, for example, primarily hold records of active journals, and journal preservation services only maintain records of participating journals. To solve this problem and to create a dataset that is as comprehensive as possible, we consulted several different data sources—title lists by the DOAJ, Ulrichsweb, Scopus title lists, and previously created datasets that might point to vanished OA journals (Björk et al., 2016; Laakso et al., 2011; Morrison et al., 2017). Except for Ulrichsweb and journals encountered during earlier research projects, we only used sources that are or have at one point been freely available on the Internet. We collected the data manually, and each data source required a unique approach for detecting potential vanished journals.

Since none of our data sources presented clear evidence for vanished journals, and only provided vague signs of potential cases at best, we needed further evidence to confirm that a journal had existed at some point and then did, in fact, vanish. We verified journals, from September–November 2019, by first checking the Keepers Registry to see if any of them were enrolled in preservation schemes as this would mean that the journals were still accessible. Since the Keepers Registry only lists digital content with an assigned ISSN, we only included journals with an individual ISSN or E-ISSN. We then tried to find the journals' websites through the ISSN database, other indexing databases, such as WorldCat, and Google searches for the journal title and ISSN. Perhaps not surprisingly, we were not able to visit the majority of journal websites with just the original web address. However, the Internet Archive's Wayback Machine once again proved to be an invaluable tool, which enabled us to access the journal websites, or most often fragments thereof, and record the year of the last published OA issue and when the journal was last available online. We define the year the journal was last available online as the year the journal vanished. Since we prioritize reliable and reproducible results over extensive yet uncertain inferences, our findings likely represent only a lower-bound estimate since identifying what no longer exists in the present poses methodological challenges. Hence, we assume that the actual number of vanished OA journals is much higher, but the currently available, fragmented data sources prevent us from verifying these with certainty. The following paragraphs elaborate on the data collection process for each of the sources we used.

With over 14,000 journals as of 2020, the DOAJ is the most comprehensive database of active OA journals. To identify vanished journals, we determined which journals have been removed from the DOAJ by cross-checking database records from 2010–2012, 2012–2014, and 2014–2019. We were able to verify that 87 journals had vanished.

Ulrichsweb Global Serials Directory offers the most extensive serial coverage with over 400,000 titles, including around 100,000 refereed scholarly journals. Similar to the approach we used for the DOAJ, we compared two title lists—one from May 24, 2012, and the other from July 3, 2018—to see which journals had been removed from the database. We were able to verify that 52 journals had vanished.

Elsevier's subscription-based database, Scopus, currently indexes over 24,000 titles (November 2019). By reviewing freely available Scopus title lists—one from February 2014 and another from April 2018—we determined which journals were no longer indexed by Scopus, and by attempting to visit their websites, we were able to verify 12 vanished journals.

One of the authors of this paper (Mikael Laakso) had also observed vanished journals during previous research projects that primarily focused on different aspects of the journal publishing landscape (Björk et al., 2016; Laakso et al., 2011). These previous studies contributed with 11 journals that we could verify as vanished. Moreover, we drew on a dataset on APC changes created by Morrison et al. (2017), which also provides information as to when journal titles could not be queried. Of these, we were able to verify 24 journals that vanished from the web. Finally, we also received a number of potential journals from two peers. Of these, we were able to verify 23 journals that vanished from the web.

## 3.2 | Creating a unified dataset

Using the data sources mentioned above, we added all these instances to a single spreadsheet to check for duplicates and merge cases that were identified by multiple data sources ($n = 34$). This resulted in 174 unique vanished journals. Of these, 154 journals have disappeared completely; seven journals have individual and incomplete issues available on the web; seven journals have print copies of some issues in libraries; and subscription services hold individual volumes and issues of six journals.

As Figure 1 shows, the individual data sources contributed mostly unique cases with only a small degree of overlap between them. The multimethod data collection strategy proved to be beneficial and only minimally redundant for vanished journals.

For the verified journals ($n = 174$), we accessed the websites through the Internet Archive's Wayback Machine to find the following information: ISSN, year founded, last year of publication, last year available online, language, country, affiliation (e.g., society, research institution), academic discipline (National Academies of Sciences, Engineering, and Medicine, 2006). Depending on the data source through which we initially identified the vanished journal, we had already acquired some of this data at previous stages. The formal data analysis is tracked with R and R Markdown and is openly available on GitHub, including version history (Jahn, 2020).

## 4 | RESULTS

### 4.1 | How many OA journals have vanished from the web?

We were able to verify 174 OA journals that have vanished from the web. In many cases, the journals first transitioned to an inactive state for several years before eventually disappearing. We want to emphasize that this should be considered as a lower-bound count and that the number of vanished journals is likely to be much greater, but identifying and verifying additional cases would require a different methodological approach.

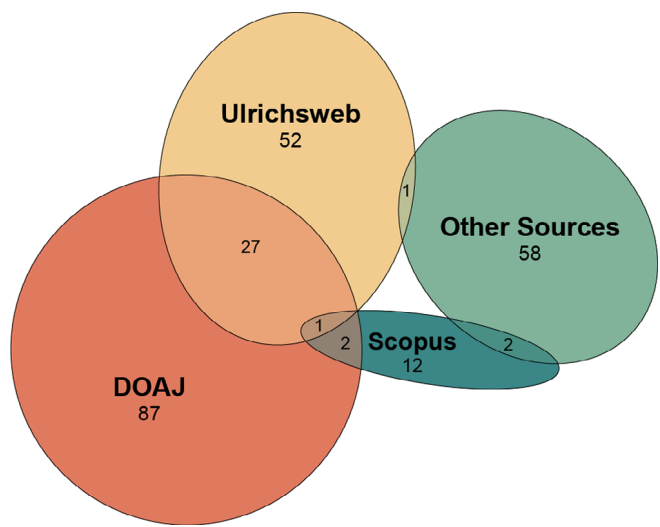

**FIGURE 1** Data source overlap and contribution to the final dataset of verified vanished journals. "Other Sources" groups vanished journals discovered by previous studies (Björk et al., 2016; Laakso et al., 2011; Morrison et al., 2017) and those submitted by our peers. The overlap between "DOAJ" and "Other Sources" ($n = 1$) is not displayed [Color figure can be viewed at wileyonlinelibrary.com]

**TABLE 2** Vanished journals categorized based on last publication year

| Last publication year | Journals | Median age | SD |
|---|---|---|---|
| 2000–2009 | 37 | 4.5 | 3.1 |
| 2010–2014 | 110 | 5.00 | 7.73 |
| 2015–2018 | 27 | 6.00 | 5.56 |
| Total | 174 | 5.00 | 6.75 |

## 4.2 | When did the journals vanish?

Our first focus point for the chronological analysis was to determine when the journals vanished from the web. Table 2 presents a numerical representation of vanished journals grouped by the year of their last publication. We found that the vast majority of journals in our sample disappeared since 2010 ($n = 137$). The 5-year span from 2010 to 2014, in particular, saw high numbers and registered more vanished journals on its own ($n = 110$) than the other two periods combined ($n = 64$). This is perhaps not surprising as we would expect journals that vanished before 2010 to be underrepresented in our sample since our data sources only date back to 2010.

However, since the year of vanishing is likely to be influenced by several factors unknown to us, we were also interested in other temporal aspects that could help us understand what kinds of journals vanish. Next, we analyzed how long the vanished journals had been publishing before becoming inactive. Figure 2 shows the publishing timeline for each of the journals. On average, the journals had been publishing for slightly over 6 years (median 5 years). Over half of the journals in our sample ceased publishing after 5 years or less ($n = 91$). However, we also encountered several cases with more extended publishing activity of 15 years or more ($n = 10$). Among these are, for example, the life sciences (LS) journal *Annales Universitatis Mariae Curie-Sklodowska – Sectio D. Medicina*, which was active between 1946 and 2010, and the *Durham Anthropological Journal*, which published between 1970 and 2013.

Finally, we were interested in how much time passed from the point the journals published their last issue to when they vanished (i.e., the last accessible, archived snapshot of the website). Figure 3 visualizes this period

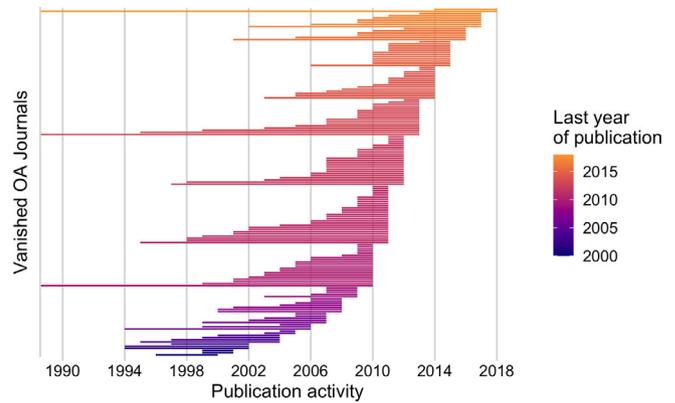

**FIGURE 2** Publication history of vanished OA journals. Each horizontal line represents an individual journal and its years of actively publishing; the journals are ordered according to the last year of publication. The line represents the period between the first and the last year of publishing. The *x*-axis is limited to the year 1990 [Color figure can be viewed at wileyonlinelibrary.com]

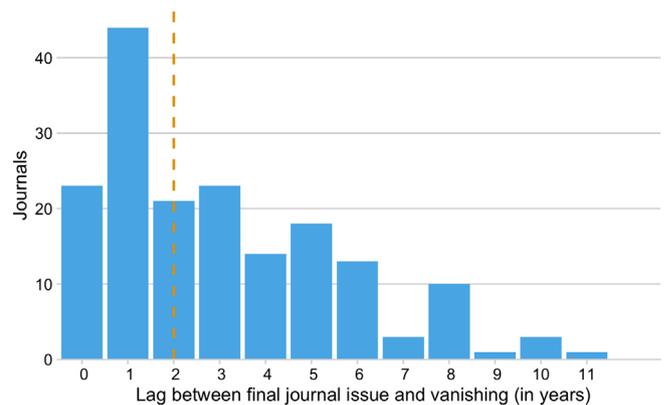

**FIGURE 3** Period between the last journal publication and vanishing in years. The vertical dashed line depicts the median [Color figure can be viewed at wileyonlinelibrary.com]

between the journals' last publication and their disappearance (in years). Over a third of the journals in our sample vanished within 1 year after the last publication ($n = 68$), and more than three-quarters had vanished within 5 years ($n = 144$). A notable exception, however, is the *African Journal of Environmental Assessment and Management*, which vanished more than 10 years after its last published issue. Table 3 provides a numerical representation of the time lag between inactivity and vanishing.

## 4.3 | What are the characteristics of vanished OA journals?

Following a general overview of the historical aspects of vanished OA journals, we wanted to zoom in on the journals' country of origin and their academic discipline. Figure 4 presents a breakdown of the journal's lifespan by academic discipline, showing only slight differences between the disciplines. Moreover, it shows that the phenomenon of vanishing journals is not limited to just one field but occurs across disciplines. Notably, social sciences and humanities (SSH) journals represent the largest share of vanished journals in our sample (52.3%), while the remaining journals are evenly split between the health sciences (Health; 16.7%), physical sciences and mathematics (PSM; 16.7%), and LSs (LS; 14.4%). However, comparing the disciplinary distribution of vanished OA journals to OA journals indexed in the DOAJ (based on Crawford, 2019) shows that SSH and PSM journals are represent a larger share of vanished than active OA journals (48.74% and 15.78% of active OA journals, respectively) whereas the opposite is true for journals from the health sciences (21.88% of active OA journals). While this comparison does not evaluate the findings statistically, it gives some indication of how the profile of the identified vanished journals compares to the profile of OA journals overall.

| Lag (in years) | Journals | Proportion (in %) | Cumulative percentage |
|---|---|---|---|
| 0 | 23 | 13 | 13 |
| 1 | 44 | 25 | 38 |
| 2 | 21 | 12 | 50 |
| 3–5 | 55 | 32 | 82 |
| >5 | 31 | 18 | 100 |
| Total | 174 | 100 | 100 |

**TABLE 3** Time lag between last active publication year and estimated year of vanishing

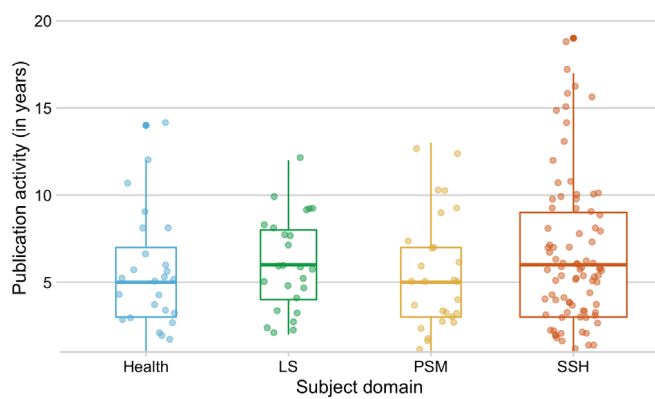

**FIGURE 4** Lifespan distribution of vanished journals across subject domains, in years. Each point represents a journal and its lifespan. The wider the box, the more vanished journals belonged to that particular discipline: Health ($n = 29$), LS ($n = 25$), PSM ($n = 29$), SSH ($n = 91$). The y-axis is limited to 20 years [Color figure can be viewed at wileyonlinelibrary.com]

Next, we analyzed the geographic distribution of the journals, finding that the 174 vanished journals in our sample were based in 47 different countries. Table 4 presents the geographic distribution of vanished journals, highlighting that this phenomenon is not just limited to one specific region but rather that it is occurring on a global scale. Nevertheless, according to our data, some regions are more affected than others. From an economic perspective, our data shows that high-income countries account for more than half of all vanished journals ($n = 107$, 61.5%; World Bank), followed by upper-middle-income countries ($n = 34$, 19.5%) and lower-middle-income countries ($n = 33$, 19%). Comparing the geographical distribution of vanished OA journals to OA journals indexed in the DOAJ (based on Crawford, 2019) reveals that North America and South Asia have a larger share of vanished than active OA journals (32.8–6.7% and 13.2–2.9%, respectively), while the opposite applies to the remaining regions. In North America and Europe and Central Asia, most vanished journals belonged to the SSH, whereas South Asia saw more LS journals disappear. As mentioned before, we followed an exploratory approach to collect data from several sources all with differing inclusion criteria and hence we did not analyze these findings statistically as this would require more systematic data collection methods.

We then looked at the journals' affiliations and found that half of the journals in our sample had an academic affiliation ($n = 86$)—either with a scholarly society ($n = 16$) or with some other kind of research organization, such as a university ($n = 70$).

Finally, Table 5 presents a breakdown of vanished journals by the journal's language relative to the publication activity in years. More than three-quarters of the vanished journals in our sample published only English articles ($n = 137$). In contrast, only slightly over one fifth of all journals also disseminated scholarly research articles in a language other than English ($n = 37$). We found no notable difference between the language of the journals and their age.

**TABLE 4**  Vanished journals grouped by major world region and academic discipline

| Region | Health | LS | PSM | SSH | Total | Proportion of vanished journals (in %) | Proportion of active OA journals (in %) |
|---|---|---|---|---|---|---|---|
| East Asia and Pacific | 5 | 2 | 6 | 5 | 18 | 10.34 | 13.91 |
| Europe and Central Asia | 8 | 3 | 8 | 33 | 52 | 29.89 | 50.96 |
| Latin America and Caribbean | 3 | 1 | 4 | 9 | 17 | 9.77 | 19.74 |
| Middle East and North Africa | 3 | 1 | 1 | 1 | 6 | 3.45 | 4.94 |
| North America | 6 | 5 | 6 | 40 | 57 | 32.76 | 6.65 |
| South Asia | 4 | 12 | 4 | 3 | 23 | 13.22 | 2.85 |
| Sub-Saharan Africa | 0 | 1 | 0 | 0 | 1 | 0.57 | 0.95 |
| Total | 29 | 25 | 29 | 91 | 174 | 100 | 100 |
| Proportion of vanished journals (in %) | 16.67 | 14.37 | 16.67 | 52.3 | 100 | | |
| Proportion of active OA journals (in %) | 21.88 | 12.71 | 16.67 | 48.74 | 100 | | |

*Note:* Academic disciplines from left to right: Health Sciences (Health), Life Sciences (LS), Physical Sciences and Mathematics (PSM), Social Sciences and Humanities (SSH). Data in the column and row listing the proportions of active OA journals is based on Crawford (2019). Crawford includes 216 journals categorized as "Other Sciences," which mostly are multidisciplinary and/or megajournals from STEM disciplines (LS and PSM in our study). To calculate the proportions of active OA journals, we divided these journals evenly into LS and PSM (i.e., 108 per group).

**TABLE 5**  Vanished journals by the journal's language relative to the publication activity in years

| Journal language | Vanished journals | Proportion (in %) | Mean age (in years) | SD |
|---|---|---|---|---|
| English | 137 | 78.74 | 6.73 | 7.25 |
| Mixed | 17 | 9.77 | 5.47 | 2.74 |
| Non-English | 20 | 11.49 | 6.45 | 5.54 |

# 5 | DISCUSSION

The primary aim of this study was to, for the first time, explore the phenomenon of vanishing OA journals. Our results indicate that OA journals are not vanishing in vast numbers, but we observed journals with certain characteristics appear more frequently than others. In particular, we find journals that were affiliated with academic institutions or scholarly societies, located in North America, or that published SSHs research, represent a larger share of vanished journals compared to other types. Finally, this study highlights the pressing lack of available data sources that track such developments consistently and comprehensively.

Presenting an initial overview, we found that only a small proportion of OA journals vanished between 2000 and 2019. However, we caution against reading this with optimism for two reasons. First, we see this as the lower estimate since the currently available data does not allow us to gauge the full extent of this phenomenon—neither do we know how many articles these journals published nor how many journals were undetected. For example, our sampling strategy excluded journals without an ISSN, yet during the data collection process we encountered several vanished journals ($n = 26$), published by Scientific Journals International, that did not have individual ISSNs but instead were all grouped together under one publisher-level ISSN. A list of identified journals without an ISSN can be found in the published dataset of inactive journals (Laakso, Matthias, & Jahn, 2020). Second, using DOAJ records as a benchmark for the total number of OA journals ($n = 14{,}068$; DOAJ, 2019), vanished journals ($n = 174$) would have only represented 1.2% of all OA journals or 2.1% ($n = 8{,}487$) of those without preservation arrangements (i.e., no information listed under "Digital archiving policy or program (s)," "Archiving: national library," or "Archiving: other" in DOAJ, 2019). An exact comparison is difficult to achieve, but the vanished journals share of 1.2% of all OA journals (i.e., currently DOAJ-indexed journals) appears relatively high compared to the much lower trigger rates of 0.02 and 0.04% for journals (OA and subscription) in Portico and CLOCKSS, respectively. While we cannot establish the reason for this discrepancy, we suspect that a majority can be attributed to the focus of preservation infrastructures on established and professional subscription-based publishers, which might be less susceptible to vanishing.

Concerning journal lifecycles we found that, on average, vanished journals actively published for slightly over 6 years and remained accessible for an additional 2 years after becoming inactive. Considering this 8-year lifecycle and that the number of OA journals has tripled over the last decade ($n = 4,767$ in 2009, $n = 14,068$ in 2019; DOAJ, 2019; Laakso et al., 2011), this might imply that a large number of OA journals is yet to vanish. Indeed, during the data collection process, we encountered almost 900 inactive OA journals that were still accessible at the time of our study but at high risk for vanishing in the near future (see Laakso et al., 2020). Moreover, our study provides valuable insight into what types of OA journals have vanished. Although university and society journals and, in particular, scholar-led journals have been at the heart of OA from the very beginning, they have also been vulnerable to financial and technical instabilities. Universities and libraries have been struggling with tight budgets (Miller, 2018; Nicholas, Rowlands, Jubb, & Jamali, 2010; Sample, 2012; Tillack, 2014), so while they champion the idea of OA, allocating the funds necessary to sustain their publishing activities and to invest in content preservation can be challenging. Our findings suggest that current approaches to digital preservation are successful in archiving content from larger journals and established publishing houses but leave behind those that are more at risk. Hence, preservation initiatives may need to re-evaluate their current strategy and develop alternative pathways—ideally in close collaboration and consultation with university and society journals—that are better suited for smaller journals that operate without the support of large, professional publishers. Recently, there have been considerable developments in this direction. Only a couple of months after we published a preprint of this study, the DOAJ, CLOCKSS Archive, Internet Archive, Keepers Registry/ISSN International Centre, and Public Knowledge Project (PKP) issued a joint press release citing the results of our study as inspiration and cause for concern, announcing collaboration on improving the preservation of small-scale, APC-free OA journals (CLOCKSS, 2020a, 2020b).

In addition to scholar-led journals, journals published in North America or the SSH represented a disproportionately large share of vanished journals in our dataset when compared to the population of active OA journals. This could point to distinct regional and disciplinary academic cultures and, indeed, a recent study found that North American SSH researchers are generally more skeptical of the benefits of OA (Bongiovani, Gómez, & Miguel, 2012; Dalton, Tenopir, & Björk, 2020). Similarly, academic career progression in North America rarely provides incentives for active involvement in OA journals (e.g., through publications or editorial roles; see Alperin et al., 2019; Niles, Schimanski, McKiernan, & Alperin, 2020). Signalizing which contributions are valued could have an impact far beyond authors' publishing decisions and also affect perceptions of what is worth preserving. The disproportionately low share of vanished journals from Latin America—where the principles of community and OA are embedded into academic culture—seems to emphasize the importance of perceived value in content preservation.

Our results suggest that vanishing of OA journals occurs across all academic disciplines and geographical regions. Furthermore, this issue should be considered as an ongoing process that will continue unless we fully commit to preserving the scholarly record. Successfully solving this issue will require the active involvement of the scholarly community as a whole and solutions as diverse as scholarly research itself. While the current system places the responsibility for preservation mainly on OA journals alone, other actors (e.g., funders, academic institutions, authors) play a vital role in facilitating this process and in mitigating losses. Over the last decade, an increasing number of research funders have implemented mandates that require beneficiaries to ensure OA to their publications by either publishing in OA journals or, when choosing subscription journals, depositing a copy of the manuscript in an OA repository (Jisc, n.d.; ROARMAP, 2018). In addition, many of these mandates also require publications to be deposited in a repository when publishing in OA journals to secure long-term access (cf. Wellcome Trust, n.d.; FWF, n.d.; European Commission Directorate-General for Research & Innovation, 2017). Academic institutions could adopt a similar approach to encourage authors to self-archive their publications, independent of the publication venue. In fact, a recent study on journal publications by faculty from around the world offers a positive outlook, finding that 81% of a sample of 620,000 OA articles are also deposited in repositories (Robinson-Garcia, Costas, & van Leeuwen, 2020). Actions like these move away from placing the sole responsibility of preserving the scholarly record on journals and toward recognizing that this responsibility is shared with all actors. Recently, coalitions S has proposed a rather radical stance on preservation, which requires authors to only publish in journals with existing preservation arrangements (Science Europe, n.d.). If implemented, such a mandate would prevent authors from publishing in the majority of OA journals indexed in the DOAJ (10,011 out of 14,068; DOAJ, 2019).

This study offers many possible avenues for future investigations into the topic of vanishing journals. From a bibliometric perspective, for example, a worthwhile topic would be how often articles from vanished journals have been and continue to be cited. Our dataset can also be used as a starting point for analyzing these journals in greater detail. Such studies could focus on what was outside the scope for this study, such as inclusion in common indexes, citation-based metrics, or potential inclusion on lists flagging journals suspected of unethical publishing practices. Moreover, future studies could draw on different data sources and in particular regional indexes, such as African Journals Online, Catálogo Latindex, or J-Stage, to detect vanished journals that might not have been indexed in the databases we used for this study (cf. Rozemblum, 2014). Researchers could also explore how to improve the monitoring of the journal landscape and detecting changes in publishing activity (e.g., inactivity or inaccessibility). This study highlights some of the consequences of the deficiencies of the current bibliometric data environment, such as inactive or vanished journals that are often delisted or removed without maintaining records of their prior existence. Bibliometric research would benefit from improved longitudinal records that follow a standardized format and track the journals' publication model and activity changes. These insights could aid the maintenance of bibliographic databases. From a sustainability perspective, surveys or interviews with editors and publishers could shed light on what causes journals to disappear and how to prevent it. Finally, the phenomenon of vanishing journals is not limited to OA but also affects digital-only subscription journals; research in this direction would also be fruitful.

Finally, we want to highlight two exceptional preservation initiatives without which our work would not have been possible. The Internet Archive and especially the Wayback Machine have proven to be invaluable resources for this research project since following the traces of vanished journals would have been much more uncertain and imprecise otherwise. In some cases, the Internet Archive also saves cached snapshots of individual articles, so they remain accessible, yet the snapshots do not necessarily amount to complete journal volumes (Ainsworth, Nelson, & Van de Sompel, 2015). Furthermore, our project greatly benefited from the work of the Keepers, which was at the brink of shutting down in 2019 due to limited financial resources. If discontinued, this would be detrimental since the Keepers is the only service monitoring preservation arrangements. Fortunately, the ISSN International

Centre has since adopted the initiative ensuring continued access.

As we have highlighted throughout the discussion, open is not forever, and so we close with a note on the urgent need for collaborative action in preserving digital resources and preventing the loss of more scholarly knowledge.

## ACKNOWLEDGMENTS

We would like to thank Richard Poynder and Dr. Ross Mounce, who pointed us toward additional vanished journals and helped us to improve this manuscript. We are grateful to the many people that reached out to us with feedback after the preprint became available, and in particular Dr. Matan Shelomi (Shelomi, 2020) and Dr. Gerta Rücker (Rücker, 2020) for publishing formal commentary documents on ArXiv.

## DATA AVAILABILITY STATEMENT

The full dataset of vanished journals is available as open data in Laakso et al. (2020). The formal data analysis is tracked with R and R Markdown and is openly available on GitHub, including version history (Jahn, 2020).

## ORCID

*Mikael Laakso*  https://orcid.org/0000-0003-3951-7990